# BciPy: Brain-Computer Interface Software in Python

Tab Memmott[I*], Aziz Koçanaoğulları[II], Matthew Lawhead[III], Daniel Klee[I], Shiran Dudy[IV], Melanie Fried-Oken[V], and Barry Oken[I]

I. Department of Neurology and School of Medicine. Oregon Health and Science University. 3181 SW Sam Jackson Rd, Portland, OR 97239. United States of America.
II. Department of Electrical and Computer Engineering. Northeastern University. 360 Huntington Ave, Boston, MA 02115. United States of America.
III. Oregon Clinical and Translational Research Institute. Oregon Health and Science University. 3181 SW Sam Jackson Rd, Portland, OR 97239. United States of America.
IV. School of Medicine. Oregon Health and Science University. 3181 SW Sam Jackson Rd, Portland, OR 97239. United States of America.
V. The Institute on Development and Disability. Oregon Health and Science University. 3181 SW Sam Jackson Rd, Portland, OR 97239. United States of America.

***Corresponding author contact information**

Full Name: Tab Memmott
Mailing Address: 3181 SW Sam Jackson Rd CR-120, Portland, OR 97239. United States of America.
Phone: 503.746.3966
Email: memmott@ohsu.edu

# Abstract

There are high technological and software demands associated with conducting brain-computer interface (BCI) research. In order to accelerate development and accessibility of BCI, it is worthwhile to focus on open-source and desired tooling. Python, a prominent computer language, has emerged as a language of choice for many research and engineering purposes. In this manuscript, we present BciPy, an open-source, Python-based software for conducting BCI research. It was developed with a focus on restoring communication using event-related potential (ERP) spelling interfaces, however it may be used for other non-spelling and non-ERP BCI paradigms. Major modules in this system include support for data acquisition, data queries, stimuli presentation, signal processing, signal viewing and modeling, language modeling, task building, and a simple Graphical User Interface (GUI).

# Introduction

Advances in software development significantly move BCI research forward. Open source software platforms such as BCI2000, BCILAB, OpenViBE, Psychtoolbox, PyFF, PsychoPy, and EEGlab have provided platforms to accelerate research in the fields of BCI and cognitive neurosciences (Brainard, 1997; Brunner et al., 2012; Delorme & Makeig, 2004; C. A. Kothe & Makeig, 2013; Peirce, 2009; Renard et al., 2010; Schalk, McFarland, Hinterberger, Birbaumer, & Wolpaw, 2004; Venthur et al., 2010) (See Table 1 for comparison of software). Despite these extant tools, there remains a need for more readily accessible software and platforms that meet end-user requirements. Preferably, new software used to develop and evaluate BCI should be openly distributed and optimized for both research and product distribution (Wessel, Gorgolewski, & Bellec, 2019). The speed and accessibility of this software is important, both for reproducibility and to help accelerate the delivery of end-user systems. We discuss two major programming languages used in the BCI field and propose Python as a superior alternative for open contribution. We will then describe our contribution to the open source effort, BciPy, originally presented at the 7$^{th}$ annual BCI Society meeting in Asilomar, California, to address the needs we highlight in this introduction (Memmott et al., 2018). This work is not meant to replace existing systems described above or outperform them; on the contrary, it is presented here as an alternate option that will integrate with existing tools wherever possible. Our primary aim is to reduce the barrier for adoption and contribution in the BCI field as a whole. This is essential for any research platform.

To illustrate potential deficiencies in the landscape of available BCI tooling resources, we cite a common BCI programming platform, MATLAB (MathWorks$^{TM}$), which utilizes a proprietary language and development environment released in 1984. The insights gained from this platform have been extensive and it remains a popular choice for instruction both inside and outside the classroom. However, as an environment, MATLAB has drawbacks which prevent it from being the ideal platform to distribute and iterate BCI software tooling. Most critically, it is not free. Individual contributors, hobbyists, smaller laboratories, and young scientists may be prohibitively discouraged from joining the research effort due to the platform's cost. Secondly, embedding MATLAB code into production level software is not an easy task, which raises concerns regarding end-user product distribution. Use of the existing distribution tools provided by MATLAB require additional financial resources and have limited scope. A final concern is speed. While the time requirements for BCI control are somewhat flexible, MATLAB is relatively slow and requires advanced configuration and multiple sessions in order to provide the same multiprocessing features offered by other languages at lower resource costs. Despite these drawbacks however, MATLAB is relatively easy to use and remains appealing to many research groups in conjunction with Psychtoolbox and MATLAB's Simulink (Brainard, 1997). To bridge this gap, new tooling should be accessible to a wide array of disciplines while maintaining the functionality and flexibility needed to research BCI.

One possible solution to the issues raised above is utilization of Python, an open-source, free, high-level programming language (Python Software Foundation) that makes use of simplified syntax and provides a plethora of out of the box tooling. These qualities make it ideal for new contributors and seasoned engineers alike. The trade-offs mentioned with using low-level languages (such as, C++) usually relate to speed and memory usage. These trade-offs are not always realized though, as many time-critical functions can have lower level bindings to increase speed and then use Python as their declarative interface. Numpy and Pandas, highly used libraries for mathematics and data science, are ideal examples of this pattern (McKinney, 2010; Oliphant & Millma, 2006; Van Der Walt, Colbert, & Varoquaux, 2011). This pattern reduces both the complexity of a given subject and also barriers of entry for scientists without programming expertise. Python has steadily increased in popularity in recent years (David Robinson, 2017) and has been adopted as the official language for universities and countrywide computer science and science courses. Furthermore, the adoption of Python in research and data science fields has led to a plethora of open source libraries and tooling options. A primary language for a BCI system that is both preferred and more easily understood by the larger field is a major benefit of BciPy over the two major adopted BCI platforms (BCI2000 and OpenVibe; See Table 1)

**Table 1: Comparison of Software for use in BCI research**

| | **Primary Language** | **Python Bindings** | **BCI focused** | **Primary BCI Modules** | **Contributions within last year** |
|---|---|---|---|---|---|
| **BciPy** | Python | yes | yes | yes | yes |
| **BCI2000** | C++ | yes | yes | yes | no |
| **OpenVibe** | C++ | yes | partial | yes | yes |
| **PsychoPy** | Python | yes | no | no | yes |
| **Psychtoolbox** | Matlab | yes | no | no | yes |
| **PyFF** | Python | yes | no | no | no |
| **BCILAB** | Matlab | yes | yes | no | yes |

**Table 1** Comparison of Software for use in BCI research. A breakdown of the primary programming language, python compatibility, focus on BCI, presence of all modules needed for BCI operation, and contributions within the last year. Many of these systems can operate on most modern operating systems, however maintaining compatibility with older systems is not guaranteed and acquisition devices may not provide drivers for all operating systems.

In this paper, we describe the structure, major functions, and theoretical considerations of the open source Python-based package, BciPy. The primary aim of the BciPy is to provide an alternative avenue for BCI research software with functionality that can be used in parallel with the various pre-existing systems. Additionally, this code demonstrates how to collect ERPs without the use of parallel port infrastructure. Historically, the high temporal precision of these ports has made them ideal for the synchronization of EEG and markers of stimuli presentation. However, new computers do not contain these ports and alternatives involve the conversion of available ports to parallel ports using other commercial hardware devices. BciPy instead uses Lab Streaming Layer (LSL), which contains callbacks that allow investigators to label stimuli in relation to continuously acquired data (C. Kothe, 2014). This method is suitable for BCI control as those labels can be used for real-time data queries.

# Methods and Materials

## Architecture

BciPy is written using object-oriented programming, a technique that models and partitions truly independent parts of an application (Figure 1). This software provides several ways to learn about each individual module, as well as the higher-level interactions necessary for BCI control. Each module contains unit tests, documentation, and demo files. The primary modules include Acquisition, Language Modeling, GUI, Signal (Processing and Modeling), Task, Display, Feedback and Helpers. With the exception of the Helpers module, the connector of various modules and convenience functions, all modules are described in greater detail below. The system timing has been validated as a whole and produces expected output for both ERP and Steady State Visual Evoked Potentials (SSVEP) (See Figure 2). This temporal fidelity is made possible by the triggers output from the Display, which may be used as query parameters to the Acquisition module. The modules are also capable of adjusting to system offsets, passed in either as arguments or inferred by reconciling differences between timestamps generated by the Display and those sent to LSL. Each module is described in more detail below.

**Figure 1: BciPy High-Level Diagram**

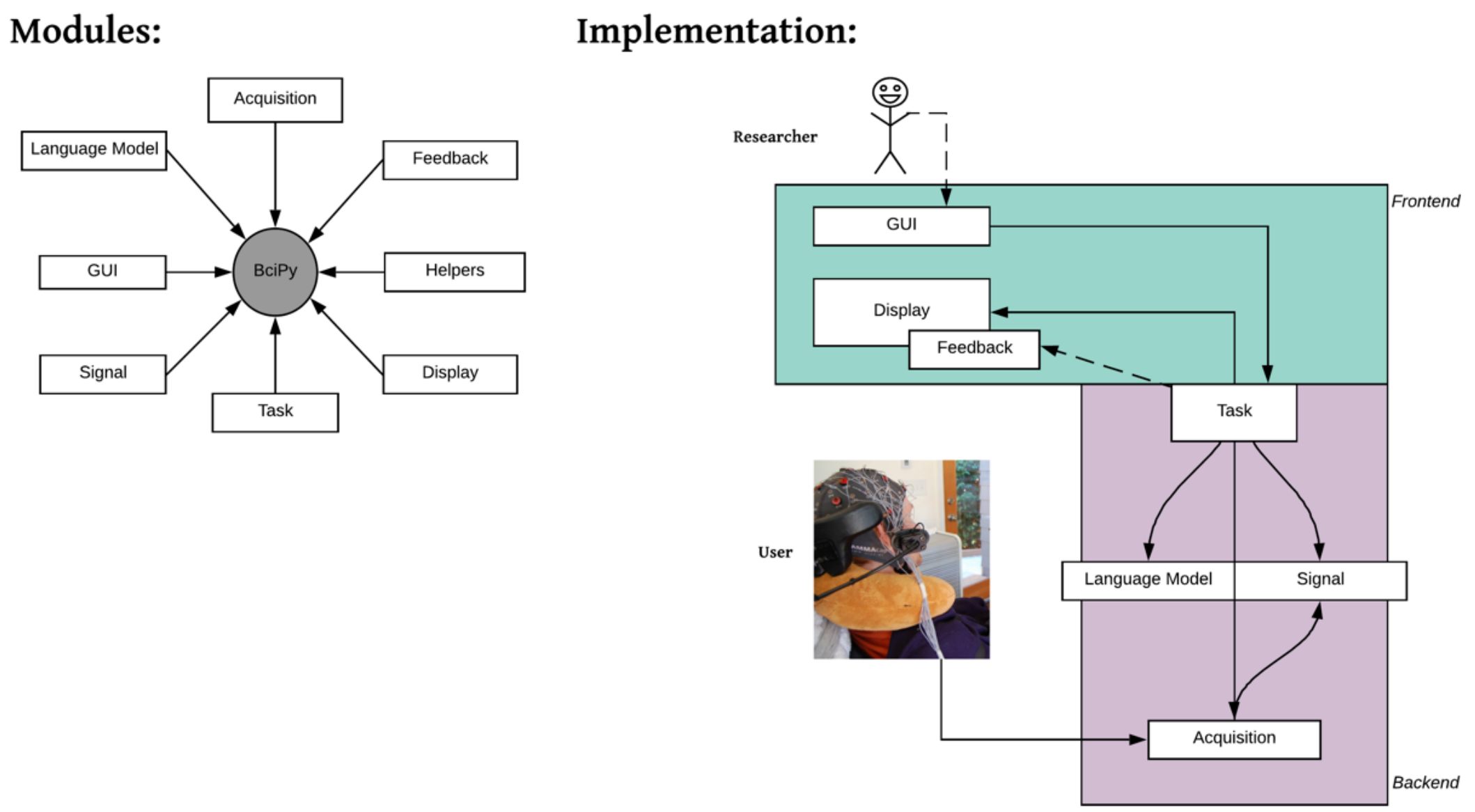

**Figure 1** BciPy high-level diagram of modules and implementation. This graphic demonstrates the necessary components for BCI use as well as how BciPy modules are partitioned in the current version. The implementation figure splits functionality into Frontend and Backend, in which the observable outputs of BciPy are considered frontward facing (Frontend) components.

**Figure 2: BciPy EEG Session Results**

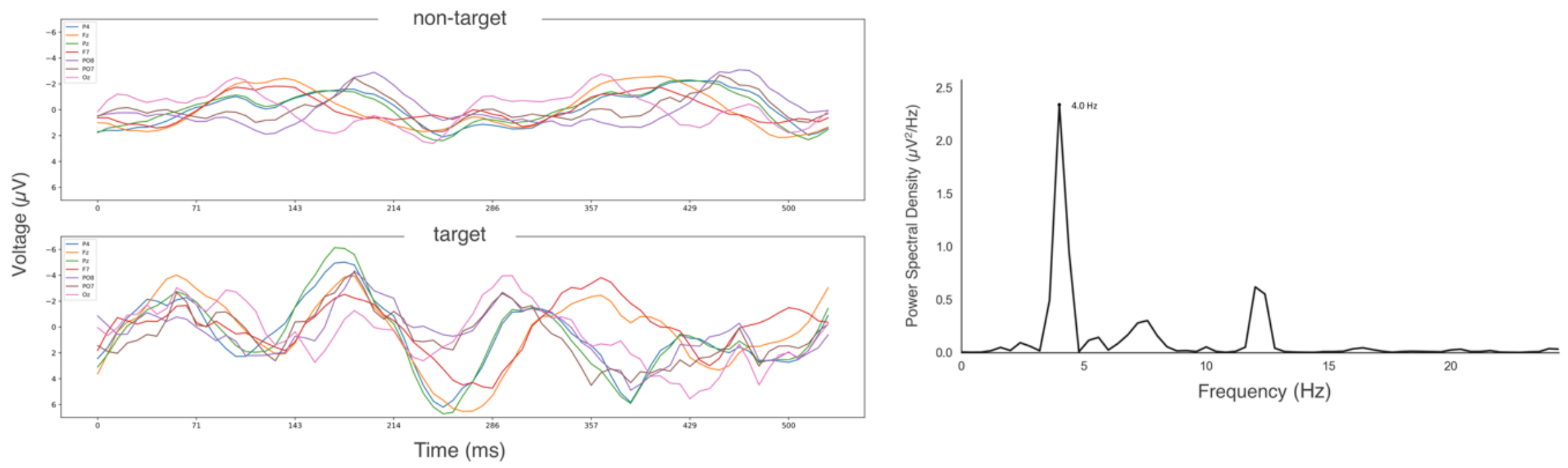

**Figure 2** EEG data collected using the BciPy system in RSVP Calibration mode collected using the Wearable Sensing's VR300 system. This demonstrates the software's ability to collect ERPs such as the P300 (left), as well as SSVEP (right) as demonstrated using PSD off the presentation rate of 4 Hz on the Oz channel.

# Acquisition

The acquisition module implements a client, which interacts with a buffer and file writer to facilitate data acquisition, writing, and real-time queries. Temporally precise real-time data queries are indispensable to BCI research.

## Architecture

The main entry point to the Acquisition module is the DataAcquisitionClient. This class is responsible for managing an incoming data stream, processing the data, and providing an interface for real-time data queries. The acquisition module makes extensive use of the multiprocessing libraries included with Python in order to avoid conflicting with the primary execution thread. DataAcquisitionClient manages two threads: one for acquisition and one for processing. The acquisition thread receives a continual stream of data and writes that data to a process queue with associated timestamps. The process thread monitors the queue and sends data to both the processor and a Buffer. Data in the buffer is archived for further queries. Each of these elements is configurable and together they provide a great deal of user flexibility.

## Hardware Interface

One of the configurable parameters associated with DataAcquisitionClient is the Device code, which DataAcquisitionClient uses for interfacing with the system hardware. BciPy currently includes two device types. The first communicates with EEG Hardware through a TCP connection. There is a specific driver for Wearable Sensing Dry Sensor Interface (DSI) headsets using this technique. BciPy also supports connection to hardware using LSL. In addition to reading EEG data, this driver also reads LSL Marker data and integrates these records for downstream processing. Supported devices can be queried dynamically through the acquisition device registry. This function is particularly useful for scripts that want to provide different command line options.

### Persistence

All data from a session are output to a .csv file for subsequent processing and analysis. In the client, this happens as part of processing using the FileWriter object. However, this detail is a configurable property of the client. Any class which implements the Processor interface can function as a surrogate. Data persistence occurs in an isolated process as a precaution so that long running processes do not block real-time data acquisition.

### Data Queries

Streaming data are sent to a Buffer object. The Buffer keeps a configurable amount of data in memory, and periodically stores that data to disk using a backend SQLite3 database. The Buffer has an interface for executing arbitrary data queries. BufferServer is the interface to the buffer and ensures that read and write actions to the buffer are asynchronous in order to protect against race conditions. These data may be accessed by giving intuitive commands to the data acquisition client described above. The ease of data retrieval is a major contribution of this library (See Figure 3).

**Figure 3: Data Acquisition Client Demo**

```
from bcipy.acquisition.client import DataAcquisitionClient
import bcipy.acquisition.protocols.registry as registry

# Query the registry for a provided device driver. In this example, we're using the DSI device.
DsiDevice = registry.find_device('DSI')
dsi_device = DsiDevice(connection_params={'host': '127.0.0.1', 'port': 8844})

# Initialize the DAQ Client
daq = DataAcquisitionClient(device=dsi_device)

# Start acquiring data
daq.start_acquisition()

# Define the query parameters in seconds.
t1 = 0
t2 = 2

# Query for data between time 1 and 2. Returns an array of time series data matching the query.
data = daq.get_data(start=time1, end=time2)

# Stop acquisition. Cleanup.
daq.stop_acquisition()
```

**Figure 3** The data acquisition client is the main BciPy interaction with any external data device. In the above code snippet, we demonstrate how to find a device in the registry, start acquisition, query for data, and stop acquisition.

### Development Tools

Running a BCI system generally requires access to specialized hardware. For development purposes, however, these resources are not always readily or practically available. The Acquisition module provides a number of developer tools that allow users to simulate various aspects of the system. Principal among these is a DataStream module, which simulates streaming EEG data. There are currently two kinds of servers provided in this module: a socket server and an LSL server. The socket server streams data through a socket connection, simulating devices that communicate over TCP. The LSL server uses the pylsl library to write to an LSL StreamOutlet. By default, these servers generate random data, but they can also stream data recorded previously by the Acquisition module. Both can be configured to interface with a data generator in order to specify data type.

**Figure 4: Data Server Demo**

```
import time

import bcipy.acquisition.datastream.generator as generator
import bcipy.acquisition.protocols.registry as registry
from bcipy.acquisition.datastream.server import DataServer

# The protocol tells the server how to provide data (binary format, etc.).
protocol = registry.default_protocol('DSI')
channel_count = len(protocol.channels)

try:
    # Start a local server on port 8844 serving random data using the DSI protocol.
    server = DataServer(protocol=protocol,
                        generator=generator.random_data,
                        gen_params={'channel_count': channel_count},
                        host='127.0.0.1', port=8844)

    # Start the server and run until a keyboard interrupt event.
    server.start()

    while True:
        time.sleep(1)

except KeyboardInterrupt:
    print('Keyboard Interrupt')
    server.stop()
```

**Figure 4** The DataServer is used to configure a data source, with its unique protocol for interaction, and serve data into BciPy. In the above code snippet, we demonstrate how to define the protocol, initialize a data server, and stop it.

The object-oriented approach adopted by the acquisition module gives it a very flexible design. Although it was developed for BciPy, the module's extensive configurability allows it to potentially be used more broadly in other data acquisition scenarios.

## Display

We implement a visual display which relies on the PsychoPy Display Window. These windows may be used with either the pyglet or pygame Python stimuli libraries, but we implement our base display using pyglet. The purpose of the Display is to present stimuli with high temporal resolution given the available hardware and return any useful timing or stimuli properties. An example of how to build the base classes for a new BCI paradigm may be observed from our porting of Rapid Serial Visual Presentation (RSVP) (Lees et al., 2018). RSVP Display is used to present stimuli on the screen in a fixed position and sequential manner. The RSVP paradigm, a P300 based speller, has been developed and evaluated by our research group using older MATLAB-based software (B. Oken, Memmott, Eddy, Wiedrick, & Fried-Oken, 2018; B. S. Oken et al., 2014; Orhan et al., 2011). The children of the Display base classes must contain a display window and, depending on the Task type, construct various stimuli and display elements vital to the experiment.

**Figure 5: Display**

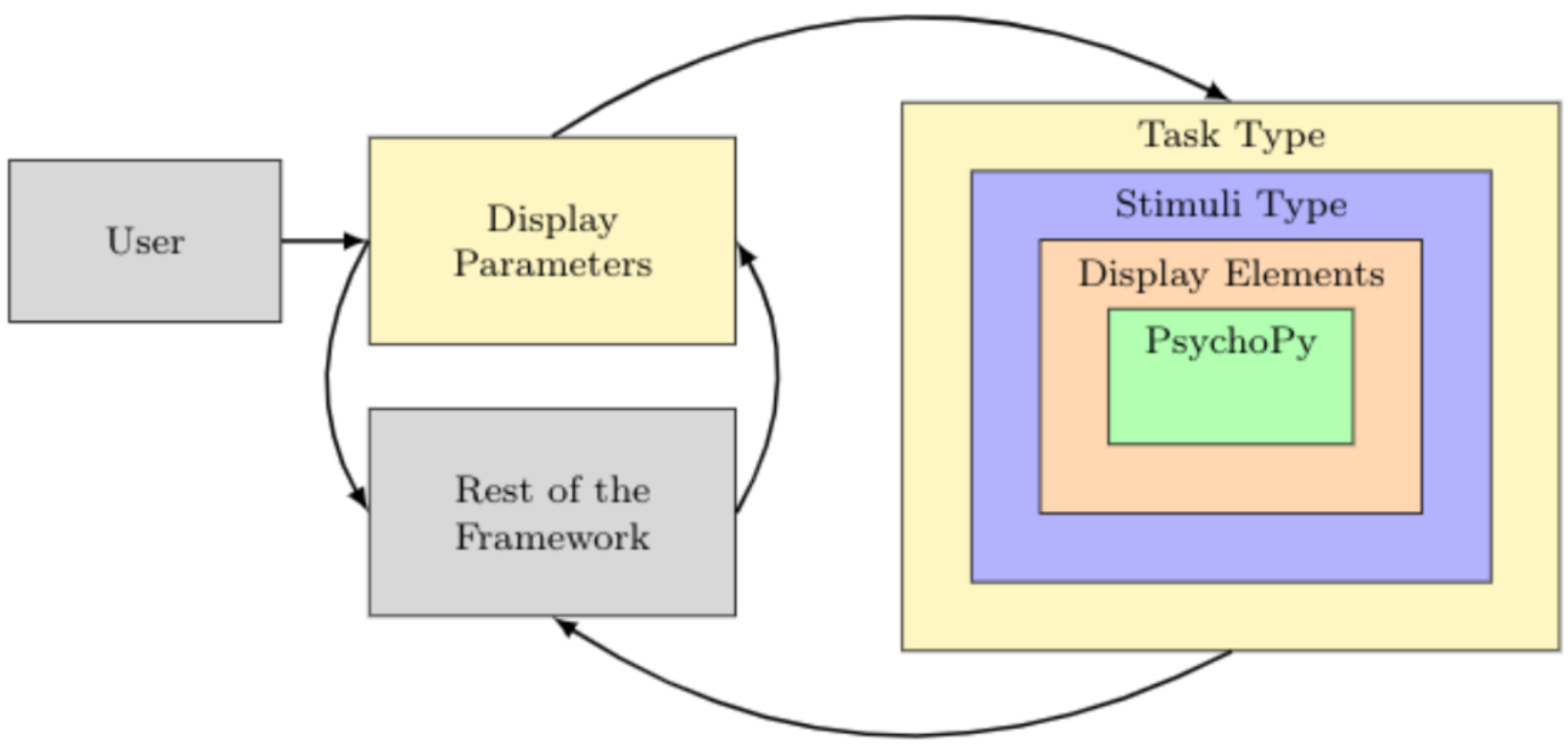

**Figure 5** The Display module diagram represents the interaction between the larger framework and the display. The parameters utilized are dependent on task type, which defines the stimuli type and display elements to be presented on the PsychoPy window.

## Graphical User Interface (GUI)

The GUI interfaces are provided for use by researchers (See Figure 1). These GUIs are meant to register modes of operation and tasks, as well as provide a method of editing registered parameters without editing those files directly. Here we describe the two interfaces currently available when using BciPy.

## BCI Interface

We provide a simple GUI that allows users to easily select registered tasks and edit operating parameters such as, presentation rate or selecting acquisition devices. It is built using WxPython and all modes of operation in BciPy are registered in this view. Currently, only the RVSP Interface is provided with, out of the box, working task implementations. However, once a paradigm is added to the task registry (See Task) it will appear in the correct GUI location. The parameter editing functionality is not dependent on task, but instead the location and correct JSON format of the parameters file.

## RSVP Interface

All RSVP tasks (see Task section below) registered in the task registry for RSVP are accessible via the GUI (Figure 6). The GUI also provides two quick link buttons for the training of EEG models after an experiment (Calculate AUC) and editing of parameters. The parameters are stored in a JavaScript Object Notation (JSON) format with value, type, recommended values, and help text information. This format allows for easy presentation, loading and saving by the user in the GUI. The supported parameter types for setting in the GUI are string, float, integer, boolean, and directory and file paths. If recommended values are provided, a dropdown will appear for easy user selection of those values. This may also prevent user's from selecting values that are not yet supported.

**Figure 6: RSVP Keyboard GUI**

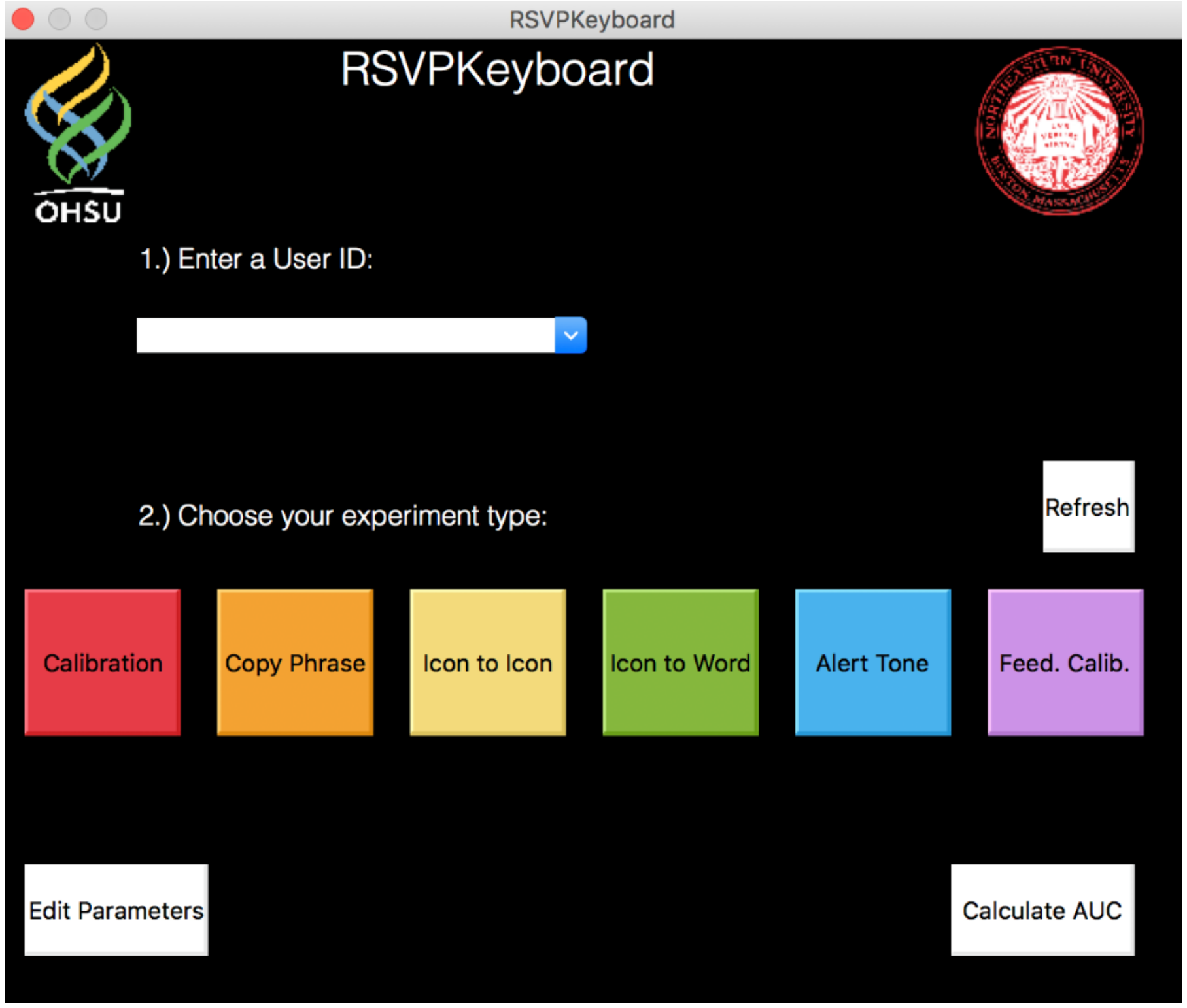

**Figure 6** The RSVPKeyboard GUI provides an easy interface to execute and configure a registered BciPy Task for RSVP. To start, a user may edit parameters, calculate AUC of a previous session, or enter a user ID. A user ID is required to start an experiment and is used to label the data folder with a timestamp for a session. After entering a user ID or selecting a previous user ID (as extracted from the data folder location in the parameters file), an experiment type may be chosen.

## Signal Viewer

BciPy GUI features an integrated Signal Viewer (See Figure 7). This GUI component allows clinicians and researchers to monitor EEG signals during an experiment to ensure that device connections are stable for consistent data quality.

**Figure 7: BciPy Signal Viewer**

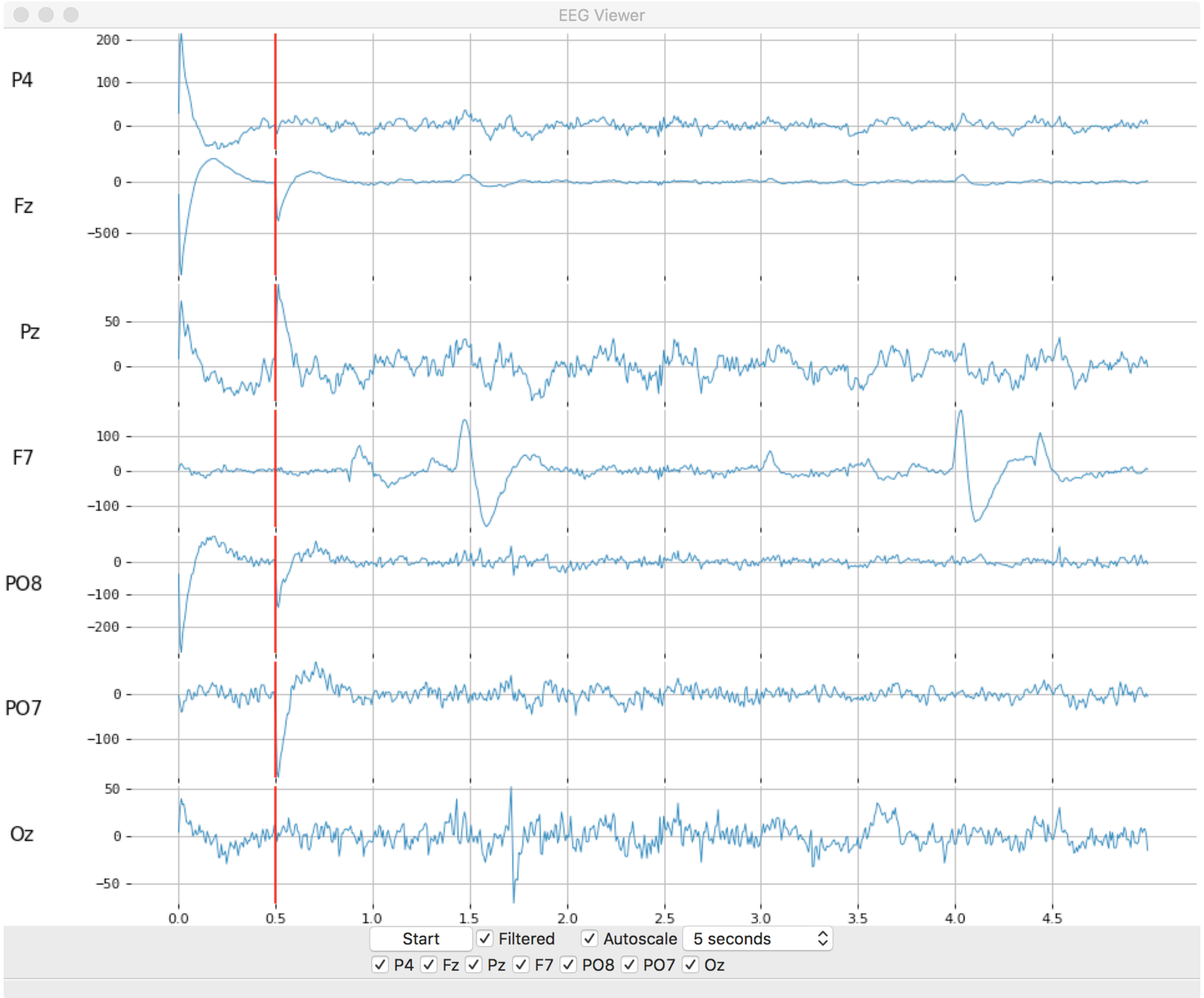

**Figure 7** The BciPy Signal Viewer displays data being served from BciPy for quality inspection before and/or during an experiment. Each available channel to BciPy is presented by default but may be removed from viewer by clicking the check boxes at the bottom with corresponding channel names. Clicking the Start button will start the stream data to the viewer that can be set at 2, 5, or 10 second update intervals. The dropdown on the bottom right allows for configuration of display window to desired length. Additionally, there are autoscaling and filtering checkboxes for convenient display configuration. The filter used in this viewer is the default bandpass filter used by all Tasks.

The Viewer is configured either through the general Configuration UI or manually in the parameters file. If selected, the Viewer will launch during the initialization of the data acquisition module in a new GUI window. The module detects the usage of multiple monitors and will appear in the secondary monitor so that it does not interfere with the main experiment. By default, all active channels are displayed. However, the Viewer has controls to toggle the visibility of any channel. Channel information is provided to the viewer, so it can work for any device supported by BciPy. There are additional controls to adjust the duration of data displayed on the screen simultaneously and to toggle filtering. Viewing can be paused at any time. Restarting from a paused state will refresh the display with the most recent data.

In addition to its use during an experiment, the Signal Viewer can be run from the command line to replay a raw data file from a previously captured BciPy session. This modality exposes some additional options for usage.

The Signal Viewer has a modular architecture, which results in a great deal of flexibility. The GUI is implemented as a WxPython Frame that streams data from any object that uses a DataSource interface. The Viewer is also parameterized with a DeviceInfo object, which provides information on sampling rate and which channels to use. Internally, the Viewer uses this metadata to determine how frequently to query any data sources for new data and to determine what channel information to expect.

Several data sources are provided in the module, including an LSL data source and a FileDatasource. The LSL DataSource is used during live experiments while the FileDataSource is used for data replay purposes. Additionally, the Viewer module integrates with the data acquisition module by implementing a custom Processor that starts the viewer in its own process. It can optionally filter this incoming data using the filters provided by the Signal module (See signal filtering section below).

Many devices come with their own proprietary software. However, use of proprietary tools often limits access to these devices, which cannot be run concurrently with other software. Other devices we encountered had outdated or non-existent software. The BciPy Signal Viewer allows users to have a consistent experience across devices to ensure good data quality.

## Language Model

Language modelling is important for the functionality of many spelling applications, including the single letter BCI RSVP paradigm where the Bayesian fusion of EEG and language model evidence improves speed and accuracy of letter typing (Orhan et al., 2011). The language model (LM) component is aimed at both improving the quality of the predictions of the system and reducing the runtime it takes to compose a message with the system. The general concept, as explained in earlier work (Manning & Schütze, 2000), is to provide information about common patterns in a language which will help direct the system towards predictions of likely symbols, given previous decisions. This section explores architecture, the types of LM modules, and how to employ them within the BciPy library.

### Architecture

The language model component acts independently of the rest of the BciPy system. The reason for this independence is to enable a “plug and play” mode that facilitates easy switching when a new LM module is developed or iteratively refined. The client's interface (found in the main script) for the two current LMs (see below) employs the same encapsulation layer, while still allowing the two LMs for minor changes in some of their functionality. The different LMs are encapsulated into a lightweight docker image (Merkel, 2014), to which queries are sent from the main BCI script. The communication between the main script (the client) and the dockerized LM module (the server) is through a TCP/IP protocol with appropriate abstraction layers on both ends.

**Figure 8: Language Model Architecture**

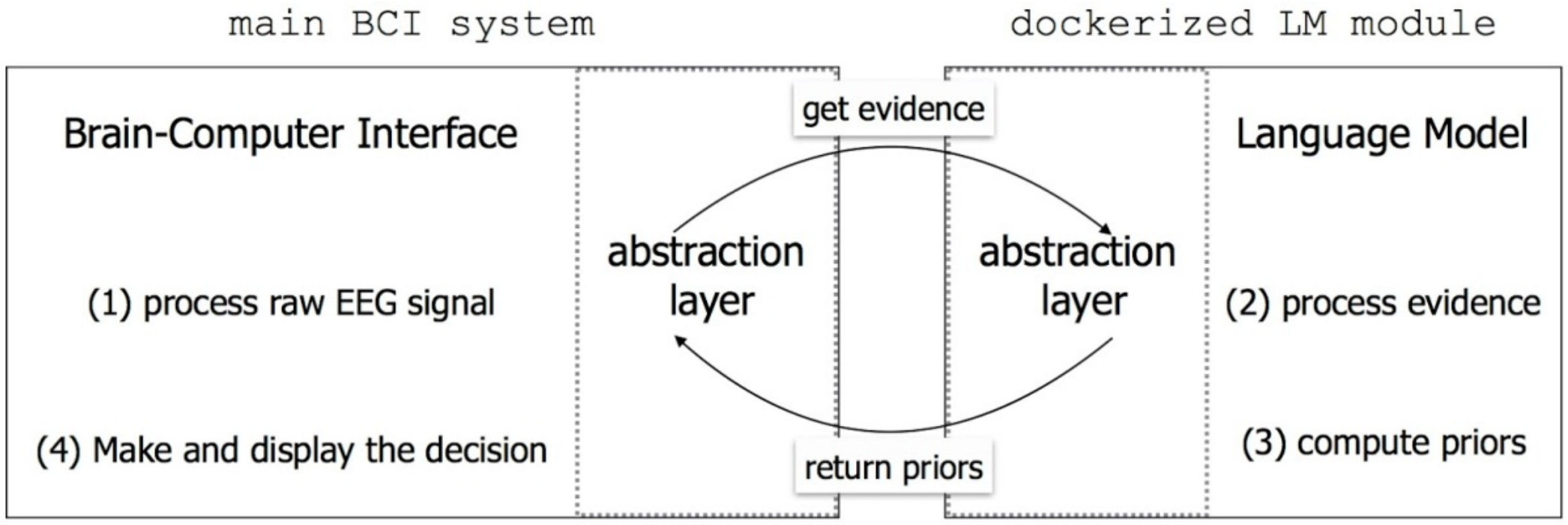

**Figure 8** Here we illustrate the process of the system from the language model perspective. First, the raw EEG is processed to enhance the underlying signal, then, it is sent over to the LM module which, along with the signal information and the language patterns information, computes a list of likely priors and sends them back to the system. Finally, the system computes the posterior distribution and makes a decision with regard to the user's symbol of intent. Technically, the communication is done through abstraction layers, containing basic functionality such as state_update, recent_priors, reset, and init, all of which are invoked by the client. In the next sections we describe the types of LM modules we have developed.

## Prefix Language Model

The concept of the Prefix Language Model (PreLM) (Dudy, Xu, Bedrick, & Smith, 2018) is to predict the next symbol (e.g., a letter, but this approach can be applied to any symbol set), given the history of symbols the user has typed. The model retrieves its prediction from a lattice that contains likely patterns of letters with associated probabilities. For example, given that the user's history is `T', the system returns a ranked list of the most likely next letters to continue the given history.

## Online Contextual Language Model

The Online Contextual LM (OCLM) is described in (Dudy et al., 2018) and is constructed to incorporate several sources of information, such as the history of likely evidence typed by the user (containing possibly more than a single letter per time step), common letter patterns, and common word paths in the language. While this approach may result in a slightly longer runtime, it provides higher quality of predictions as described in the previous paper.

### Interface

The system allows the user to define an LM module of choice, which is then propagated through the rest of the system. As a result, the user only needs to ensure that the LM image is found on their computer. Internally, the major commands remain the same for both module types, since both go through the same encapsulation layer on the client's end that communicates with the docker image, as mentioned previously.

OCLM can provide not only a ranked list of letters, but also a ranked list of words that the user is currently in the middle of typing. For both modules, the letters and the words are sent back to the client with their corresponding probabilities.

### Usage

Demonstration scripts for both model types are found in the language model module in a demo directory.

In the PreLM demo code shown in Figure 9, the following steps are taken: instantiating an LM object; getting priors for no history; getting priors to accumulated history; and resetting the LM object to delete the history of the model and enable a fresh start without regenerating the object, which would require generating a new docker instance.

**Figure 9: Prefix Language Model Demo**

```python
from bcipy.language_model.prelm_language_model import LangModel

# Initialize Language Model
lmodel = LangModel(logfile='lmlog.log')
lmodel.init()

# Get initial priors
lmodel.recent_priors()

# Get priors given letters typed
lmodel.state_update(['T'])  # priors given 'T' being typed
lmodel.reset()

lmodel.state_update(['TH'])  # priors given 'TH' being typed
lmodel.reset()

lmodel.state_update(['THE'])  # priors given 'THE' being typed
```

**Figure 9** An example of how to initialize and retrieve inferences from the Prefix Language Model.

The OCLM demo code in Figure 10 demonstrates similar steps: instantiating an LM object; determining n-best candidate; getting priors for no history; (not presented generating evidence); sending evidence along with the type of desired output (letter/word) to the object to get updated priors; and finally, a resetting step, similar to resetting in the PreLM.

**Figure 10: Online Contextual Language Model Demo**

```
from bcipy.language_model.oclm_language_model import LangModel

# Initialize Language Model
lmodel = LangModel(logfile='lmlog.log')
lmodel.init()

# Get initial priors
lmodel.recent_priors()

# Get priors given evidence and letter return mode
return_mode = 'letter'
lmodel.state_update(evidence, return_mode)

# Get priors given evidence and word return mode
return_mode = 'word'
lmodel.state_update(evidence, return_mode)
```

**Figure 10** An example of how to initialize and retrieve inferences from the Online Contextual Language Model.

# Signal

## Signal Process

The signal process module contains data filters and decomposition functions. These processing modules are necessary for making use of EEG signals for BCI control.

### Filtering

Several data filters are available for use. The primary method, bandpass, implements a Butterworth Filter that accepts arguments for high/low cutoff and filter order. Additionally, a text filter representing filter parameters created in a legacy MATLAB version of the system was implemented to retain backwards compatibility with the older, verified RSVP Keyboard system (Orhan et al., 2012). The interface for interacting with these filters is simple and optimized for use with EEG (See Figure 11). While we provide these base filtering implementations, other filters may be added and leveraged.

**Figure 11: Signal Filtering Demo**

```
from bcipy.signal.generator.generator import gen_random_data
from bcipy.signal.process.filter import bandpass, downsample, notch

# Set data parameters
raw_data = gen_random_data(low=-1000, high=1000, channel_count=25)
sampling_rate = 300

# Set filter parameters
notch_filter_frequency = 60
filter_high = 50
filter_low = 2
filter_order = 2
downsample_rate = 2

# Filter it!
notch_filterted_data = notch.notch_filter(raw_data, sampling_rate, notch_filter_frequency)
bandpass_filtered_data = bandpass.butter_bandpass_filter(
    notch_filterted_data,
    filter_low,
    filter_high,
    sampling_rate,
    order=fitler_order)
final_data = downsample.downsample(bandpass_filtered_data, factor=downsample_rate)
```

**Figure 11** Data filtering is a critical part of BCI use. In the above snippet, we demonstrate a common data filtering pipeline. We start by generating some random data and then define our sampling rate and filter parameters. We then call on the notch and bandpass filter, and lastly downsample our data which returns final_data for usage.

### Decomposition

The decomposition module implements downsampling and a power spectral density (PSD) method, allowing for extraction of discrete frequency bands, plotting of spectra, and relative power band calculations. Two decomposition method types are available for use in the current version: Welch's and Multitaper. To use the PSD function, raw data, frequency band of interest (e.g., 1-10Hz), data sampling rate, data window length (e.g., 500ms), and PSD method type are required as arguments. Consumers of this method may set a Boolean of plot=True to visualize the resultant PSD output during experiment building. The default is set to False, as plotting would be distracting during real-time operation. Additionally, a Boolean flag may be passed to return PSD as a relative calculation to the full spectrum.

## Signal Modeling

In our system, user intent is detected in a Bayesian fashion that performs posterior probability updates. The updates require a likelihood assessment between the stimuli and the user intent. We use $\sigma, \phi, \varepsilon$ to denote user intent, stimuli and EEG evidence respectively. After each stimulus presentation the system updates the belief over all possible $\sigma$'s as the following;

$$p(\sigma|\phi, \varepsilon) = p(\sigma)\frac{p(\varepsilon|\sigma, \phi)}{p(\varepsilon|\phi)} \tag{1}$$

To update the belief over the user intent, the system searches for the presence of an event-related potential (ERP), an anomaly in the EEG sequence. Each sequence contains multiple symbols being presented and hence results in a list of (multichannel) evidence and a list of corresponding labels. In the sequence, each stimulus letter has a fixed-known location and hence one can assign a label to each position. For simplicity, within this section we use positional argument $\mathrm{t}$ over the stimuli. We denote the evidence set with $\varepsilon = \{\varepsilon_1, \varepsilon_2, \cdots, \varepsilon_T\}$ as $\varepsilon_t \in \mathbb{R}^{C\times N}$ (where $\mathrm{C}, \mathrm{N}$ denotes number of channels and signal length respectively), stimuli set with $\phi = \{\phi_1, \phi_2, \cdots, \phi_T\}$ and the label sequence with $\ell = \{\ell_1, \ell_2, \cdots, \ell_T\}$ where $\ell_t \in \{0,1\}$. The position $\mathrm{t}$ has the corresponding label is $\ell_t = \delta_{\phi_t,\sigma}$; The label becomes 1 where the letter presented at location $\mathrm{t}$ is the user intent and 0 otherwise. Therefore, $\ell_t = 1$ coincides with the ERP presence. This notation allows us to write the following;

$$p(\varepsilon_t|\sigma, \phi_t) = \begin{cases} p(\varepsilon_t|\ell = 1) & if\ \sigma = \phi_t \\ p(\varepsilon_t|\ell = 0) & otherwise \end{cases}$$

Signal modelling is responsible for generating the likelihoods, in words the likelihood of evidence resulting from the label sequence $p(\varepsilon|\ell)$. In our application we assume each evidence chunk is independent of each other conditioned on the labels. This allows us to simplify the update defined in (1) for estimates that appear in the stimuli as the following;

$$p(\sigma|\phi, \varepsilon) \propto p(\sigma)\frac{p(\varepsilon_t|\ell = 1)}{p(\varepsilon_t|\ell = 0)}\ where\ \phi_t = \sigma \tag{2}$$

We refer reader to our previous work for more details (Kocanaogullari, Erdogmus, & Akcakaya, 2018).

In this application, the signal modelling is tasked to return likelihoods for evidence $p(\varepsilon_t|\ell_t = 1)$, $p(\varepsilon_t|\ell_t = 0)$, conditioned on positive and negative classes respectively. In order to extract that information, we propose a pipeline of models which we explain in 3 sub-sections; Feature extraction, dimensionality reduction and generative model. We rely on scikit-learn library (Pedregosa Fabian et al., 2011) in our implementation and hence follow the conventions of the package. Let $\mathrm{x}, \mathrm{y}$ denote data and label sets respectively. Each module to be presented includes the following properties:

- fit(x, y): given a data and label sequence, fits the model
- transform(x): given a data sequence, returns transformation based on fitted model
- fit_transform(x, y): given a data and label sequence, fits the model and returns transformation based on the fitted model.

## Feature extraction

In order to extract features, principal components along each channel data are obtained with a channel wise principal component analysis (PCA) approach. To extract the components, a channel specific PCA matrix $A_{c_{M_c \times N}}$ is learned for channel c using the corresponding evidence set $\{\varepsilon_1^c, \varepsilon_2^c, \cdots, \varepsilon_T^c\}$ of the channel. Each channel PCA matrix further used to reduce the dimensionality of the corresponding channel c. To incorporate all channel information at a position t, the feature vector is formed as the following;

$$f_t = \left[A_1 \varepsilon_t^1, A_2 \varepsilon_t^2, \ldots, A_C \varepsilon_t^C\right]^T \in \mathbb{R}^d$$

Observe that $d = \sum_c M_c$. For simplicity of the application, this feature vector is assumed to be Gaussian. The simplest representation can be represented using the quadratic decision boundary between such feature vectors and we discuss the built-in method in the following section.

## Dimensionality reduction

Feature vector is passed through a regularized discriminant analysis (RDA) to obtain one dimensional representation. The model is specifically tailored to improve the intent selection by discriminating target and non-target representations. We refer the reader for further reading to Friedman's work (Friedman, 1989). RDA modifies quadratic decision boundary between classes using two hyperparameters; $\lambda \in [0,1]$, $\gamma \in [0,1]$ for shrinkage and regularization. Shrinkage balances class conditioned covariances with the overall mean subtracted covariance. As $\lambda \rightarrow 1$, both classes share the covariance and hence decision boundary becomes linear. As $\gamma \rightarrow 1$ both conditional covariances converge to circular covariance. Each of hyper-parameters are selected with cross validating the area under receiver operating characteristics curve (AUC=AUroC).

RDA returns a log-likelihood $\log\big(p(\ell = 1|f_t)\big) \in \mathbb{R}$, $\log(p(\ell = 0|f_t)) \in \mathbb{R}$ and we obtain a log-likelihood ratio score $s_t = \log(p(\ell = 1|f_t)) - \log(p(\ell = 0|f_t))$ number for each trial where $s_t \in \mathbb{R}$. By design, $s_t > 0$ if the stimuli is the user intent and $s_t < 0$ otherwise.

## Generative Model

To perform the posterior update presented in (2) we stated that we need $p(\varepsilon|\ell)$, which is calculated along class conditional likelihood distributions $p(\varepsilon_t|\ell_t)$. We estimate these likelihoods from a training data where label assignment is already known using kernel density estimates (KDEs) with Gaussian kernels around each log-likelihood score point $s_t$. These models are tasked to provide the conditional likelihoods $p(\varepsilon_t|\ell_t = 1)$, $p(\varepsilon_t|\ell_t = 0)$ during the inference for the update presented in (2).

**Figure 12: Signal Model Processing Pipeline**

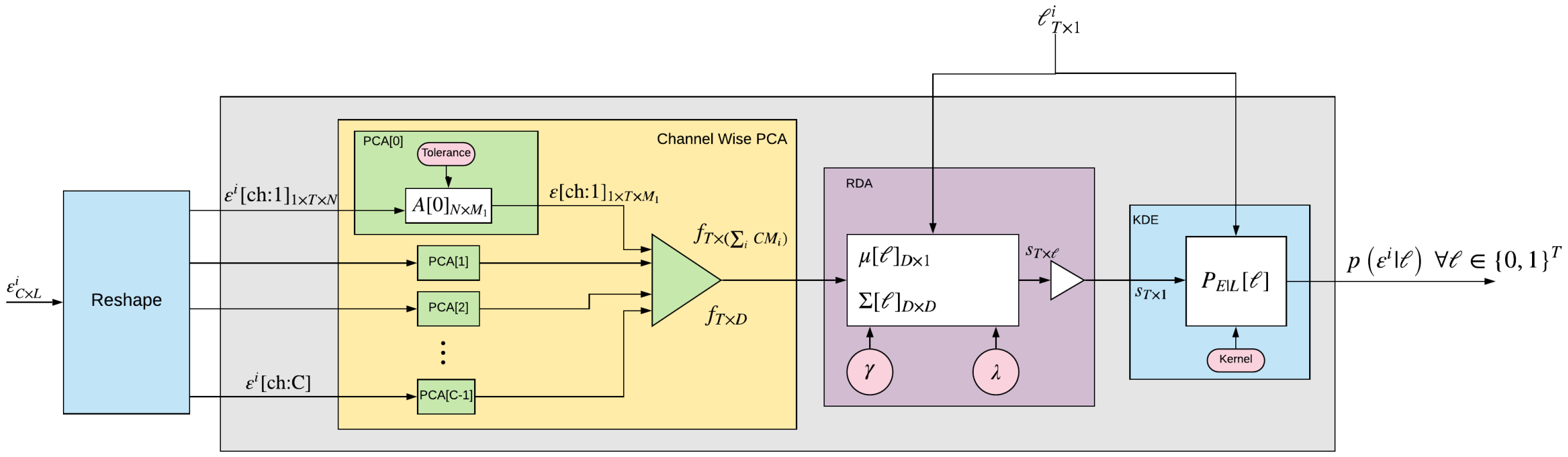

**Figure 12** Signal Processing pipeline for likelihood generation. At a high level, reshaped data is sent throught the pipeline to undergo Channel-Wise PCA, RDA, and finally KDE. The output is then returned to the system to make decisions on.

<u>Use-case:</u> The pipeline requires the evidence ε and label $\ell$ to train model parameters. The pipeline is visualized in Figure 12. Each model in the pipeline can be used independently by their own methods. To give an example, the code in Figure 13 defines RDA, fits a mean and a covariance with fixed $\lambda \in [0,1]$, $\gamma \in [0,1]$ and outputs the transform, defined as z.

**Figure 13: Signal Modeling Regularized Discriminant Analysis Demo**

```
from bcipy.signal.model.mach_learning.classifier.function_classifier import RegularizedDiscriminantAnalysis

rda = RegularizedDiscriminantAnalysis()
z = rda.fit_transform(x, y)
```

**Figure 13** An example of how to initialize the BciPy Regularized Discriminant Analysis class and use it to transform data, where x is reshaped data and y are labels.

Hyperparameters are set using the cross validation. The cross validation optimizes hyperparameters to maximize area under receiver operation characteristics curve for all the validation sets. Cross validation accepts x, y as training samples, the entire model and element to be optimized (See Figure 14).

**Figure 14: Signal Modeling Cross Validation Demo**

```
from bcipy.signal.model.mach_learning.cross_validation import cross_validation

cross_validation(x, y, model, opt_el, k_folds=10, split='uniform')
```

**Figure 14** An example of how to initialize the BciPy Cross Validation class and use it to validate inferences. The inputs are as follows: x is reshaped data, y is the data labels, model is a trained model, and opt_el is element to optimize on. A user may also optionally specify the number of folds and define the split type.

In the proposed pipeline presented, the element to be optimized is RDA with required parameters which is the first element.

Each of the functionalities are presented with their own demo files within the demo folder inside 'machine learning (mach_learning)' sub directory.

# Feedback

The feedback module provides functionality for both auditory and visual feedback types. The module may be used in a Task and inherit a Display to present stimuli. However, it may also be used independently. Visual feedback uses PsychoPy visual and core modules to administer and configure visual feedback stimuli. Users can build on top of visual feedback to administer progress, alertness, attention, and other context-based feedback during an experiment without complicating their main display or task logic. Additionally, using the Python library python-sounddevice, built on top of PortAudio (Bencina & Burk, 2001), auditory feedback may also be administered. The RSVPInterSequenceFeedbackCalibration task is an example of visual feedback being used in conjugation with EEG frequency activity to focus attention during a calibration session (Klee, McLaughlin, Memmott, Fried-Oken, & Oken, 2019).

# Task

BciPy implements its own Task module, which is the basis for experimental design and where the modules connect to form operational meaning. A Task at its base level must contain implementation specific settings, a unique name, and a method of execution. All tasks programmed into the task registry are accessible via the GUI. These tasks can be used to accomplish the closed-loop BCI control (See Figure 15) needed for many BCI experiments. Currently, tasks to calibrate, spell, and provide feedback in an RSVP paradigm (Orhan et al., 2012) are included with the base classes for utilization. For example, the RSVP copy phrase task utilizes all modules to collect, query, and save data, retrieve language and signal inference, display stimuli and ultimately provide spelling ability.

**Figure 15: Closed Loop BciPy Control**

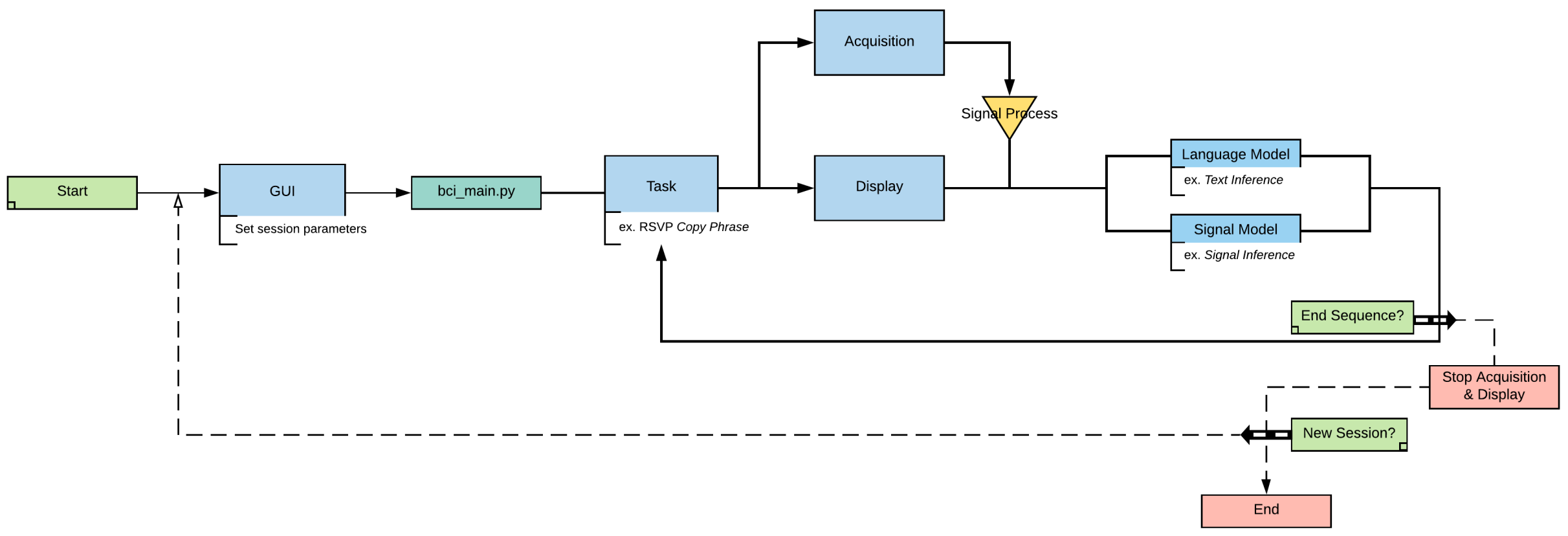

**Figure 15** Closed loop control diagram of BciPy. The blue and yellow entities indicate BciPy modules, and green/red boxes are user interaction or criteria checks. Starting at the leftmost side, a user starts a BciPy GUI (such as RSVPKeyboard.py), selects experimental parameters, and begins a session via the main entry point bci_main.py at the root of the project. After passing requisite parameters to bci_main.py, the Task starts with Acquisition and Display running in parallel processes. After a presentation sequence, data is piped through signal processing and modeling, and (optionally) Language Modeling to decide on what to do next. The Task, having implemented some decision-making criteria, then decides the next presentation sequence or to discontinue sequencing. Each loop results in written data, checking of criteria, and formation of new stimuli or ending the session. If stopping criteria

are met, Acquisition and Display are stopped in their parallel processes, data are written, and the user is directed back to the GUI.

## System Requirements

The system currently requires a graphics card capable of OpenGl support and an operating system which complies with the libraries used by BciPy, particularly PsychoPy. BciPy uses only Python 3, as support for Python 2 is reaching end of support from the Python Software Foundation in January 2020. Currently, the system can install on most modern Windows and Mac OS. Manual installation of some dependencies would be required for Linux installation. New computers should be tested for timing before formal experimentation and blocking applications (such as, some Antivirus software) configured to cease during operation. Additionally, configuring new systems, especially laptops, may require disabling power save mode which may default to onboard graphics instead of performance graphics card needed for timing resolution in order to save battery. To this purpose, there are scripts provided in the repository for visualizing the offset of triggers and raw data, as well as some documentation on how to remedy common issues, either in real time or offline. These resources, along with traditional photodiode assessments should be used when starting new experiments or preparing new computers for use.

## Release and Contribution

BciPy is hosted on PyPi and pip installable (Python Community, 2014). The code is hosted on GitHub and can be accessed using the following link: https://github.com/BciPy/BciPy. To use the provided RSVP implementation, the user will need to git clone or download the repository and follow the README instructions for local usage. We invite all contributors to the repository and will enforce the Code of Conduct, utilizing the Contributor Covenant v1.4.1, listed at the root of the repository to encourage a safe development environment. It is currently distributed under a BSD license. Please refer to the LICENSE.txt in the repository for more information.

# Discussion

BciPy moves the state of BCI research forward by providing functionally dynamic development tools in Python. The technological requirements for BCI research are immense, and it is essential that accessible tooling be obtainable to the larger community. While our implementations may not suit all BCI applications and many facets of the field are quickly accelerating, our tools have been designed to handle additions and new functionalities as they become available. These early releases are meant to lay the foundations needed for all BCI research with some implementations as examples as they are developed. This should make adoption for paradigms not currently implemented lower than if starting from scratch or using some of the software described in Table 1. For instance, we’ve presented many tools to operate event related BCI paradigms, particularly for spelling, however more work would need to be done for robotic control. From an accessibility and software engineering perspective, future contributions that bridge existing research software will be particularly beneficial to future releases. Moreover, improvements that allow for greater OS compatibility and easier calibration of stimuli timing are needed. There remain hurdles in setting up new commercial machines to be performant for BCI use.

Additionally, tools needed to test timing at the start of an experiment, such as a photodiode, may be difficult to operate if not included in a commercial set. There is ongoing work to release debug and visualization tools to help diagnose timing and other critical aspects to BCI operation.

## Acknowledgments

We utilized many learnings from our group's original MATLAB implementation of RSVP Keyboard (Orhan et al., 2012). We thank David Smith, Shaobin Xu, Deniz Erdogmus, Steven Bedrick, Brandon Eddy, Betts Peters, Deirdre McLaughlin, Ian Jackson, and Dani Smektala for their architectural considerations, code and feedback over the course of development. Funding was provided by NIH grant #R01 DC009834 and the National Institute on Disability, Independent Living, and Rehabilitation Research (NIDILRR grant #90RE507).

## Declaration of Interest

The authors whose names are listed on this manuscript have no conflicts of interest to report.